\documentstyle[aaspp4,tighten,flushrt]{article}

\newcommand{\beq}{\begin{equation}}
\newcommand{\eeq}{\end{equation}}
\newcommand{\beqa}{\begin{eqnarray}}
\newcommand{\eeqa}{\end{eqnarray}}

\newcommand{\lexp}{\mathop{\langle}}
\newcommand{\rexp}{\mathop{\rangle}}
\newcommand{\rexpc}{\mathop{\rangle_c}}

\def\d{\delta}

\font\BFd=cmmib10
\font\BFt=cmmib10
\font\BFs=cmmib10 scaled 700
\font\BFss=cmmib10 scaled 500

\def\bbox#1{%
\relax\ifmmode
\mathchoice
{{\hbox{\BFd #1}}}
{{\hbox{\BFt #1}}}
{{\hbox{\BFs #1}}}
{{\hbox{\BFss #1}}}
\else \mbox{#1} \fi }

\def\d{\delta}

\def\ds{\delta_s}

\def\del{\nabla}

\def\ga{\mathrel{\mathpalette\fun >}}
\def\fun#1#2{\lower3.6pt\vbox{\baselineskip0pt\lineskip.9pt
        \ialign{$\mathsurround=0pt#1\hfill##\hfil$\crcr#2\crcr\sim\crcr}}}

\def\k{{\bbox{k}}}
\def\x{{\bbox{x}}}
\def\r{{\bbox{r}}}
\def\s{{\bbox{s}}}

\def\q{{\bbox{q}}}
\def\uz{{\hbox{$u_z$}}}
\def\u{{\bbox{u}}}
\def\v{{\bbox{v}}}

\begin{document}

\title{Gravitational Clustering from $\chi^2$ Initial Conditions}

\vskip 1pc

\author{Rom\'{a}n Scoccimarro}

\vskip 2pc

\affil{Institute for Advanced Study, School of Natural Sciences,
Olden Lane, Princeton, NJ 08540}

\begin{abstract}

We consider gravitational clustering from primoridal non-Gaussian
fluctuations provided by a $\chi^2$ model, as motivated by some models
of inflation.  The emphasis is in signatures that can be used to
constrain this type of models from large-scale structure galaxy
surveys. Non-Gaussian initial conditions provide additional non-linear
couplings otherwise forbidden by symmetry that cause non-linear
gravitational corrections to become important at larger scales than in
the Gaussian case. In fact, the lack of hierarchical scaling in the
initial conditions is partially restored by gravitational evolution at
scales $k> 0.1$ h/Mpc.  However, the bispectrum shows much larger
amplitude and residual scale dependence not present in evolution from
Gaussian initial conditions that can be used to test this model
against observations.  We include the effects of biasing and redshift
distortions essential to compare this model with galaxy redshift
surveys. We also discuss the effects of primordial non-Gaussianity on
the redshift-space power spectrum and show that it changes the shape
of the quadrupole to monopole ratio through non-linear corrections to
infall velocities.

\end{abstract}

\subjectheadings{large-scale structure of universe}

\clearpage 

\section{Introduction}

The current paradigm for the formation of large-scale structures in
the universe is that small primordial fluctuations, with a roughly
scale-invariant spectrum, are amplified by non-linear gravitational
instability to form the galaxies, clusters and superclsuters that we
see today in galaxy surveys.  A common assumption is that the
statistics of these primordial fluctuations is Gaussian, which is
consistent with current observations.  However, at this point we
cannot rule out the possibility that primordial fluctuations were in
fact non-Gaussian to some extent.  Fortunately, rapid progress in
microwave background anisotropy experiments and large galaxy redshift
surveys in the next few years will provide high quality data that can
be used to test the nature of primordial fluctuations.

The consideration of non-Gaussian initial conditions is a difficult
issue because there is an infinite class of non-Gaussian models.  In a
Gaussian field $\phi$, $n-$point connected correlation functions,
$\xi_n = \lexp \phi_{1} \ldots \phi_{n} \rexp_{c}$ of order $n \geq 3$
vanish; however, in a non-Gaussian field higher-order correlations can
virtually behave arbitrarily subject to the constraint of
realizability, that is, that the multivariate probability
distributions be positive definite.  In general, however, we can
divide non-Gaussian models into {\em weakly} and {\em strongly}
non-Gaussian, depending on the magnitude of normalized cumulants
$s_{p} \equiv \lexp \d^p \rexpc / \lexp \d^2 \rexp^{p/2}$ of the
smoothed density contrast $\d$ compared to unity.  A minimal way of
achieving strongly non-Gaussian models is by {\em dimensional
scaling}, where the hierarchy of $n$-point correlation functions obeys
$\xi_n \propto \xi_2^{n/2}$.  In this case, normalized cumulants are
numbers not necessarily smaller than unity.  This is opossed to {\em
weakly non-Gaussian models} where normalized cumulants are forced to
be small quantities, e.g.  {\em hierarchical scaling} models, where
$\xi_n \propto \xi_2^{n-1}$ with $\xi_2 \ll 1$.

In this paper we consider $\chi^2$ initial conditions, which belong to
the class of dimensional scaling models. As a strongly non-Gaussian
model, it has the potential to be more easily constrained or detected
by large-scale structure observations than primordial weakly
non-Gaussian models. Furthermore, there is a number of inflationary
models in the literature that motivate $\chi^2$ initial conditions
(e.g. Kofman et al. 1989; Antoniadis et al. 1997, Linde \& Muhanov
1997; Peebles 1997).  In addition, this model has been recently argued
to fit a significant set of observational constraints (Peebles 1999a,
1999b). It is also possible that this particular model may be a good
representation of the general behavior of dimensional scaling models,
and thus provide general insight about their advantages and
disadvantages. Furthermore, models in which primordial fluctuations
are generated by topological defects also generally obey dimensional
scaling (e.g. Turok \& Spergel 1991).

Previous work on clustering from non-Gaussian initial conditions was
done using numerical simulations (e.g. Moscardini et al. 1991;
Weinberg and Cole 1992; Coles et al. 1993; White 1999), and
perturbation theory (Fry \& Scherrer 1994; Jaffe 1994; Chodorowski \&
Bouchet 1996; Gazta\~naga \& Fosalba 1998). In this work, we
concentrate on aspects of $\chi^2$ initial conditions that can be
tested with current and future observations in galaxy redshift
surveys. In particular, we concentrate on the study of the
redshift-space power spectrum and bispectrum; applications of these
results to observations will be considered elsewhere (Scoccimarro et
al. 2000).

Different aspects of the effects of primordial non-Gaussianity on the
power spectrum have been considered in the literature (e.g. Feldman,
Kaiser \& Peacock 1994; Stirling \& Peacock 1996; Sutherland et
al. 1999). The skewness of the smoothed density field in texture
models was studied by Gazta\~naga \& M\"ah\"onen (1996), whereas the
impact of hierarchical scaling models of primordial non-Gaussianity on
the bispectrum was considered by Verde et al. (1999). Our approach is
complementary to recent studies of the impact of primordial
non-Gaussian models in other aspects of large-scale structure
(e.g. Robinson, Gawiser \& Silk 1999; Koyama, Soda \& Taruya 1999;
Willick 1999; Pierpaoli, Garc\'{\i}a-Bellido \& Borgani 1999).

This paper is organized as follows. In Section~2 we review
perturbation theory results regarding the evolution of the bispectrum
from general initial conditions. Section~3 presents results for the
particular case of $\chi^2$ initial conditions. In Section~4 we
discuss the effects of galaxy biasing and Section~5 those of redshift
distortions. Finally, we present our conclusions in Section~6.

\section{Gravitational Clustering from Non-Gaussian Initial Conditions}

We are interested in the effects of non-Gaussianity on clustering
statistics at large scales. A convenient approach is to use non-linear
perturbation theory (PT), where the density field in Fourier space
at time $t$ reads (Fry 1984)

\beq
\d(\k,t) = \d^{(1)}(\k,t) + \d^{(2)}(\k,t) + \ldots 
= D_{1} \d_I(\k) + D_{1}^2 \int \d_D(\k-\k_{12}) \
F_2(\k_1,\k_2) \d_I(\k_1) \d_I(\k_2)+ \ldots,
\label{den2}
\eeq

\noindent where $\d_I(\k)$ is the initial density contrast,
$k_{i\ldots j} \equiv \k_i + \ldots + \k_j$, and we assumed that to a
very good approximation, in second-order PT fluctuations grow
according to $D_{2}(t) \propto D_{1}^2(t)$, with $D_{1}(t)$ the linear
growth factor. The kernel $F_2(\k_1,\k_2) \equiv 5/7 + 1/2 \cos\theta
(k_1/k_2+k_2/k_1) + 2/7 \cos^2 \theta$ with $\cos\theta \equiv \k_1
\cdot \k_2 /(k_1 k_2)$, describes to leading order the non-local
evolution of the density field due to the long-range nature of
gravitational interactions.

The non-Gaussianity of initial conditions is encoded in the
statistical properties of the random field $\d_I(\k)$, in particular,
its lowest order connected moments are

\beqa
\lexp \d_I(\k) \rexp &=& 0 ,\\
\lexp \d_I(\k_1) \d_I(\k_2) \rexp &=& \d_D(\k_{12})\ P^I(\k_1), \\
\lexp \d_I(\k_1) \d_I(\k_2) \d_I(\k_3) \rexp &=& \d_D(\k_{123})\ B^I(\k_1,\k_2,\k_3), \\
\lexp \d_I(\k_1) \d_I(\k_2) \d_I(\k_3) \d_I(\k_4)  \rexpc &=& \d_D(\k_{1234})\ T^I(\k_1,\k_2,\k_3,\k_4),
\eeqa

\noindent where $P^I$, $B^I$ and $T^I$ denote the power spectrum,
bispectrum, and trispectrum of the initial density field. In linear
PT, these just scale with the linear growth factor, but at the scales
of current galaxy surveys, non-linear corrections can be
significant. From Eq.~(\ref{den2}) we derive non-linear corrections to
the power spectrum and bispectrum 
\begin{eqnarray}
P(k) &=& P^I(k) + 2 \int d^3 q\ F_2(\k+\q,-\q)\ B^I(\k,\q),
\label{pknl} \\
B_{123} &=& B_{123}^I + B_{123}^G + B_{123}^T,
\label{Bispnl}
\end{eqnarray}

\noindent where $B_{123}^I$ denotes the contribution of the initial
bispectrum scaled to the present time using linear PT,
$B_{123}^I(t)\propto [D_{1}(t)]^{3}$, $B_{123}^G$ represents the usual
gravitationally induced bispectrum from Gaussian initial conditions
(Fry 1984)

\begin{equation}
B_{123}^G = 2 F_2(\k_1,\k_2)\ P^I(k_1) P^I(k_2) + {\rm cyc.},
\label{BG} 
\end{equation}

\noindent and the last term in Eq.~(\ref{Bispnl}) 

\begin{equation}
B_{123}^T = \int d^3q\
F_2(\k_{12}-\q,\q)\ T^I(\k_1,\k_2,\k_{12}-\q,\q) + {\rm cyc.},
\label{BT} 
\end{equation}

\noindent represents the contribution coming from the initial
trispectrum linearly evolved to the present. Note that only the first
term in Eq.~(\ref{Bispnl}) scales as $[D_{1}(t)]^{3}$, the last two
terms have the same scaling with time, $[D_{1}(t)]^{4}$, and therefore
eventually dominate at late times.

\section{Evolution from $\chi^2$ Initial Conditions}

We now evaluate the results of the previous section for the particular
case of $\chi^2$ initial conditions. In this case, the density field
after inflation is proportional to the square of a Gaussian scalar
field $\phi(\x)$, $\rho(\x) \propto \phi(\x)^2$. Thus, the density
constrast $\d(\x) = \phi(\x)^2 /\sigma^2_\phi -1$, where
$\sigma^2_\phi \equiv \lexp \phi^2 \rexp $. If the auxiliary Gaussian
field has two-point correlation function $\xi_\phi \equiv \lexp \phi_1
\phi_2 \rexp$, then the density 2-, 3-, and 4-point correlation
functions are respectively (Peebles 1999b)

\beqa
\xi^I_{12} &=& 2 \frac{\xi_\phi^2(r_{12})}{\sigma_\phi^4},
\label{xiI}\\
\zeta_{123}^I &=& 2^{3/2} \sqrt{ \xi_{12}^I \xi_{23}^I \xi_{31}^I},
\label{zetaI}\\
\eta_{1234}^I &=& 4 \left[ \sqrt{ \xi_{12}^I \xi_{23}^I \xi_{34}^I
\xi_{41}^I } + \sqrt{ \xi_{12}^I \xi_{24}^I \xi_{43}^I
\xi_{31}^I } + \sqrt{ \xi_{13}^I \xi_{32}^I \xi_{24}^I
\xi_{41}^I } \right],
\label{etaI}
\eeqa

\noindent and the initial density power spectrum, bispectrum, and
trispectrum are given by

\beqa
P^I &=& 2 \int d^3 q\ P_{\phi}(q) P_{\phi}(|\k-\q|),
\label{powerI}\\ 
B^I_{123} &=& 8 \int d^3 q\ P_{\phi}(q)P_{\phi}(|\k_1-\q|)
P_{\phi}(|\k_2+\q|),  
\label{bispI}\\
T^I_{1234} &=& 16 \int d^3 q\ P_{\phi}(q)P_{\phi}(|\k_1-\q|)
P_{\phi}(|\k_{14}-\q|) P_{\phi}(|\k_{3}+\q|) + 
16 \int d^3 q\ P_{\phi}(q)P_{\phi}(|\k_1+\q|)
\nonumber \\
& & \times P_{\phi}(|\k_{23}-\q|) P_{\phi}(|\k_{2}-\q|)  
+ 16 \int d^3 q\ P_{\phi}(q)P_{\phi}(|\k_1+\q|)
P_{\phi}(|\k_{24}-\q|) P_{\phi}(|\k_{2}-\q|) ,  
\label{trispI}
\eeqa

\noindent where $\sigma^2_\phi P_{\phi}(k)$ denotes the power spectrum
of the $\phi$ field. For scale-free spectra, $P_{\phi}(k) = A
k^{n_\phi}$, 

\beq
P^I(k) = 2 \pi^{3/2} \frac{ \Gamma^2
\left(\frac{n_\phi+3}{2}\right) \Gamma
\left(-n_\phi-\frac{3}{2}\right)}{\Gamma^2 (-n_\phi/2)
\Gamma(3-n_\phi)} \ A^2 k^{2n_\phi+3},
\label{PI}
\eeq

\noindent similarly the bispectrum can be expressed in terms of
hypergeometric functions, using the results in Scoccimarro (1997).
Simple analytic results can be obtained for the particular case
$n_{\phi}=-2$, which gives a density spectral index $n=-1$, reasonably
close to the observed one at the non-linear scale. Using
Eq.~(\ref{pknl}) we get 

\begin{equation}
P(k) = \frac{2 \pi^3 A^2}{k} + \frac{96 \pi^4 A^3}{7},\ \ \ \ \ \ \ \
\ \ B^I_{123} = \frac{8 \pi^3 A^3}{k_1 k_2 k_3}.
\label{powbispI} 
\end{equation}

\noindent However, the initial trispectrum does not seem to have a
simple closed form. Defining the non-linear scale from the linear
power spectrum, $\Delta^I(k_{nl}) \equiv 4\pi k_{nl}^3 P^I(k_{nl})
\equiv 1$, we have

\begin{equation}
\Delta(k) = \left( \frac{k}{k_{nl}} \right)^2\ \left( 1 +
\frac{24}{7\sqrt{2} \pi} \frac{k}{k_{nl}} \right) \approx \left(
\frac{k}{k_{nl}} \right)^2\ \left( 1 + 0.77 \frac{k}{k_{nl}} \right),
\label{delnl} 
\end{equation}

Rather than working with the bispectrum itself, it is convenient to
consider the reduced bispectrum $Q_{123}$ defined by

\beq
Q_{123} = \frac{B_{123}}{\Sigma_{123}}
\equiv \frac{B_{123}}{P_1 P_2 +P_2 P_3 +P_3 P_1},
\label{Q123}
\eeq

\noindent which for Gaussian initial conditions has the important
properties that it is independent of time, and to a very good
approximation independent of the matter density parameter $\Omega$. In
addition, for scale-free initial conditions is independent of overall
scale, for CDM-type models the scale-dependence is only weak through
the scale variation of the spectral index. From these results, the
reduced bispectrum for $\chi^2$ initial conditions, including
non-linear gravitational corrections is

\begin{eqnarray}
Q_{123} &=& \frac{4\sqrt{2}}{\pi}\ \frac{k_{nl}}{k_1+ k_2+ k_3}-
\frac{192}{7\pi^2} \frac{k_1 k_2+k_2 k_3+k_3 k_1}{(k_1+ k_2+ k_3)^2}
+ Q_{123}^G + Q_{123}^T,  
\label{QI} 
\end{eqnarray}

\noindent where $Q_{123}^G$ denotes the reduced bispectrum obtained
from Gaussian initial conditions, and $Q_{123}^T$ denotes the
contribution from Eq.~(\ref{BT}) which is difficult to calculate
analytically (however, we shall give a full numerical evaluation of
the bispectrum below). In particular, for equilateral configurations

\beq
Q_{eq}=\frac{4\sqrt{2}}{3 \pi}\ \frac{k_{nl}}{k} - \frac{64}{7 \pi^2}
+ \frac{4}{7} + Q^T_{eq} \approx 0.6 \frac{k_{nl}}{k} -0.35 + Q^T_{eq},
\label{Qeq}
\eeq

\noindent where we used that $Q^G_{eq}=4/7$ (Fry 1984). Since the last
term is a number independent of scale, Eq.~(\ref{Qeq}) illustrates the
signature of this type of non-Gaussian initial conditions: $Q_{eq}$
shows a strong scale dependence at large scales as $k/k_{nl}
\rightarrow 0$. This is not just a property of $\chi^2$ initial
conditions, but rather of dimensional scaling models ($\xi_n \propto
\xi_2^{n/2}$). 

A simple generalization of $\chi^2$ initial conditions is to consider
$N$ independent fields, each of them $\chi^2$ distributed with the
same power spectrum $P_{\phi}(k)$. As $N$ increases, the Gaussian
limit is achieved as a result of the central limit theorem. This might
be a useful way of parametrizing constraints on non-Gaussianity from
large-scale structure observations. For a fixed linear density power
spectrum $P^I(k)$, the primordial bispectrum and trispectrum obey $B^I
\propto 1/\sqrt{N}$ and $T^I \propto 1/N$,
respectively. Equation~(\ref{QI}) then reduces to

\begin{eqnarray}
Q_{123} &=& \frac{1}{\sqrt{N}} \frac{4\sqrt{2}}{\pi}\
\frac{k_{nl}}{k_1+ k_2+ k_3}- \frac{1}{N} \frac{192}{7\pi^2} \frac{k_1
k_2+k_2 k_3+k_3 k_1}{(k_1+ k_2+ k_3)^2} + Q_{123}^G + \frac{1}{N}
Q_{123}^T.
\label{QIN} 
\end{eqnarray}

\noindent Thus, the approach rate to the Gaussian initial conditions
result is $1/\sqrt{N}$ at large $N$. The same scaling holds for the
skewness parameter, similarly, the kurtosis relaxes towards the
Gaussian initial conditions value as $S_4 \propto 1/N$. Note, however,
that the scaling at small $N$ is stronger; in fact, such a behavior is
seen in the N-body simulation results by White (1999) on the skewness
and kurtosis as a function of $N$ (see his Fig.~7).

From Eq.~(\ref{delnl}) we see that non-linear corrections to linear PT
can be significant even at wavenumbers smaller than the non-linear
scale. In order to check for significant contributions from third and
higher-order in the perturbation expansion [Eq.(\ref{den2})], we
resort to numerical realizations of second-order Lagrangian PT (2LPT),
which by being formulated in Lagrangian space incorporate the
remaining terms in Eq.(\ref{den2}), although only approximately beyond
$F_2$ (but the error is small, see e.g. Scoccimarro (1998) for a
quantitative comparison of one-point cumulants).  In this case, the
perturbation expansion is done about particle trayectories $\x(t)$,

\beq 
\x(t) = \q + \Psi(\q,t) = \q + \Psi^{(1)}(\q,t) + \Psi^{(2)}(\q,t)
+\ldots,
\label{lag}
\eeq

\noindent so that $\q$ represents the initial (Lagrangian) position of
a particle whose current (Eulerian) position is $\x(t)$ and
$\Psi(\q,t)$ denotes the displacement vector assumed to be a small
quantity. In 2LPT, Eq.~(\ref{lag}) is truncated at second order. The
solutions for the displacement field are obtained from the equations
of motion and read

\beqa
\x(\q) &=& \q -D_1\ \del_q \phi^{(1)} + D_2\ \del_q \phi^{(2)},
\label{dis2} \\
\del_q^2 \phi^{(1)}(\q) &=&  \d(\q), 
\label{phi1} \\
\del_q^2 \phi^{(2)}(\q) &=& \sum_{i>j} [\phi_{,ii}^{(1)}(\q)\
\phi_{,jj}^{(1)}(\q) - (\phi_{,ij}^{(1)}(\q))^2],
\label{phi2}
\eeqa

\noindent where to a very good approximation, $D_2 \approx -3D_1^2/7$
(Bouchet et al. 1995). For a detailed exposition of 2LPT see
e.g. Buchert et al. (1994) and Bouchet et al. (1995), who also
compared to N-body simulations. Similarly, in a companion paper
(Scoccimarro 2000), we explore the validity of 2LPT for the evolution
of the bispectrum in redshift-space from Gaussian initial conditions,
by comparing to N-body simulations. In this paper, we use 2LPT for
$\chi^2$ initial conditions and adopt the criterion of validity found
for the Gaussian case, namely, we only include waves up to a maximum
wavenumber $k_c \approx 0.5$ h/Mpc, so that the percentage of
shell-crossing is below $10\%$. Given that 2LPT is computationally
inexpensive, we can generate many realizations which is essential to
beat down cosmic variance.

\begin{figure}[t!]
\centering
\centerline{\epsfxsize=18truecm\epsfysize=18truecm\epsfbox{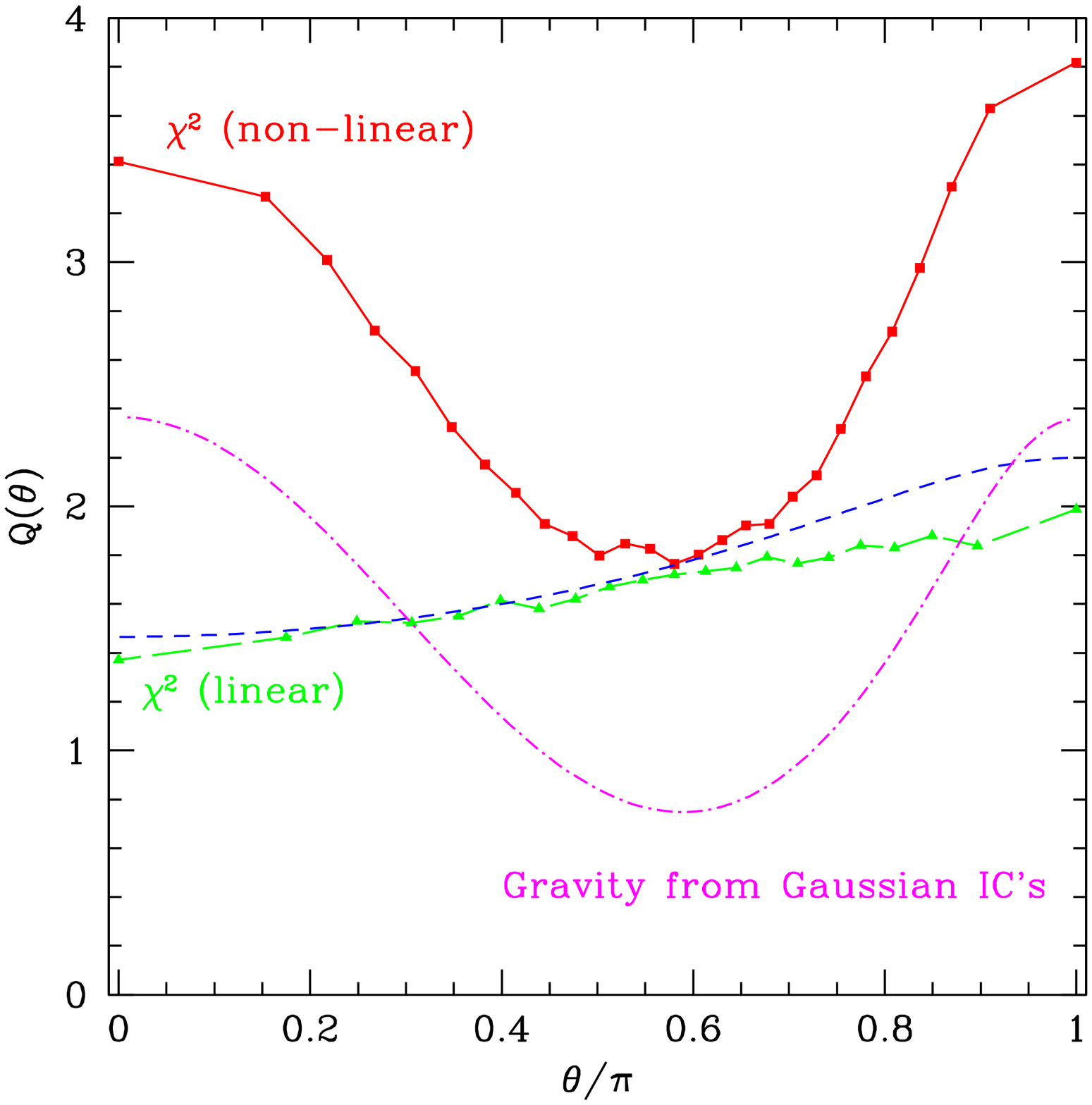}}
\caption{The reduced bispectrum $Q$ for triangles with sides
$k_1=0.068$ h/Mpc and $k_2=2 k_1$ as a function of the angle $\theta$
between $\k_1$ and $\k_2$. Triangles denote linear extrapolation from
$\chi^2$ initial conditions, whereas square symbols show the result
of non-linear evolution. Dot-dashed lines show the predictions of
non-linear PT from Gaussian initial conditions with the same initial
power spectrum as the $\chi^2$ model.}
\label{figiso}
\end{figure}

\begin{figure}[t!]
\centering
\centerline{\epsfxsize=18truecm\epsfysize=18truecm\epsfbox{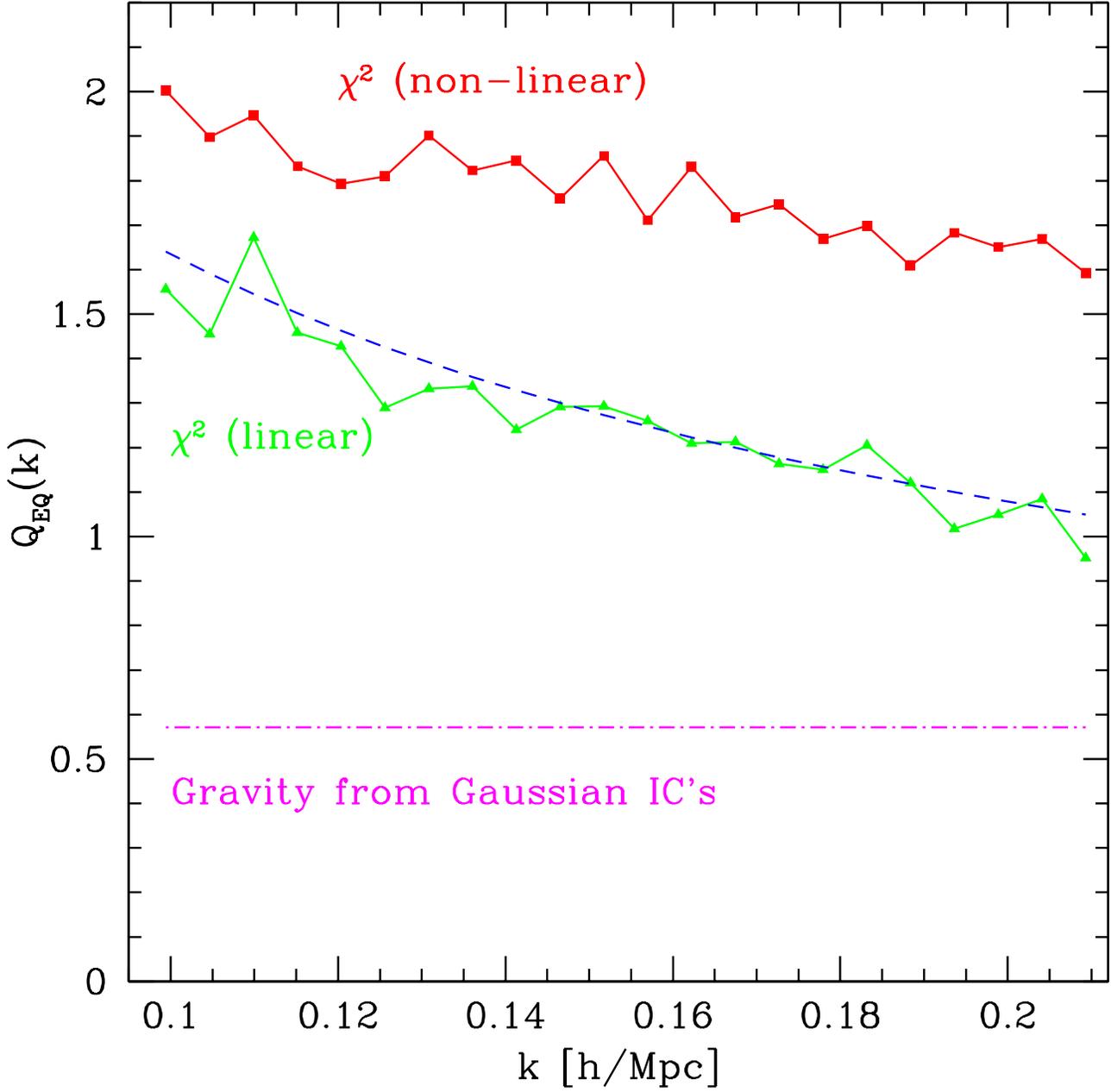}}
\caption{The reduced bispectrum for equilateral triangles as a
function of scale $k$. Line styles as in Fig.~\protect\ref{figiso}. }
\label{figisoeq}
\end{figure}

Figures~\ref{figiso} and \ref{figisoeq} show the results of 100 2LPT
realizations. We have chosen the auxiliary Gaussian field $\phi$ with
a spectral index $n_{\phi}=-2.4$, leading to $n=-1.8$ as proposed in
Peebles (1999a). The amplitude of the power spectrum has been chosen
to give $k_{nl} \equiv 0.33$ h/Mpc. First, we checked that the initial
conditions were generated correctly by comparing the power spectrum
and the bispectrum with theoretical expectations. The dashed lines in
Fig.~\ref{figiso} show the predictions of the first term in
Eq.~(\ref{QI}) for the reduced bispectrum at $k_1=0.068$ h/Mpc, $k_2=2
k_1$, as a function of angle $\theta$ between $\k_1$ and $\k_2$.  This
corresponds to $n=-1$, however, it approximately matches the numerical
results (triangles, $n=-1.8$). The latter show less dependence on
angle, as expected because the scale dependence in the $n=-1.8$ case
($Q^I \propto k^{-0.6}$) is weaker than for $n=-1$ ($Q^I\propto
k^{-1}$). In Fig.~\ref{figisoeq} we show equilateral configurations as
a function of scale for $\chi^2$ initial conditions (triangles) and
$Q^I_{eq}(k) = 0.8 (k/k_{nl})^{-0.6}$ (dashed lines), where the
proportionality constant was chosen to fit the numerical result, this
is slightly larger than the prediction in the first term of
Eq.~(\ref{Qeq}) for $n=-1$, and closer to the real-space result
$Q_{eq}(r) = 0.94 (r/r_{nl})^{0.6}$.

The behavior of the $\chi^2$ bispectrum is notoriously different from
that generated by gravity from Gaussian initial conditions for
identical power spectrum (dot-dashed lines in
Figs.~\ref{figiso}-\ref{figisoeq}) (Frieman \& Gazta\~naga 1999). The
structures generated by squaring a Gaussian field roughly correspond
to the underlying Gaussian high-peaks which are mostly spherical, thus
the reduced bispectrum is approximately flat. In fact, the increase of
$Q^I$ as $\theta \rightarrow \pi$ seen in Fig.\ref{figisoeq} is
basically due to the scale dependence of $Q^I$, i.e. as $\theta
\rightarrow \pi$, the side $k_3$ decreases and thus $Q^I$ increases.

As shown in Eqs.~(\ref{QI}-\ref{Qeq}), non-linear corrections to the
bispectrum are significant at the scales of interest, so linear
extrapolation of the initial bispectrum is insufficient to make
comparison with current observations. The square symbols in
Figs.~\ref{figiso} and \ref{figisoeq} show the reduced bispectrum
after non-linear corrections are included. As a result, the familiar
dependence of $Q_{123}$ on the triangle shape due to the dynamics of
large-scale structures is recovered (Fig.~\ref{figiso}), and the scale
dependence shown by $Q^I$ is now reduced (Fig.~\ref{Qeq}). However,
the differences between the Gaussian and $\chi^2$ case are very
obvious: the $\chi^2$ evolved bispectrum has an amplitude about 2-4
times larger than that of an initially Gaussian field with the same
power spectrum. Furthermore, the $\chi^2$ case shows residual scale
dependence that reflects the dimensional scaling of the initial
conditions. The analogous results for the skewness were obtained using
numerical simulations (White 1999) and non-linear PT in the spherical
collapse approximation (Gazta\~naga \& Fosalba 1998).  These
signatures can be used to test this model against obervations;
however, before we can do so we have to test the robustness of these
conclusions against the effects of galaxy biasing and redshift
distortions.

\section{Galaxy Biasing}

We now consider the effects of local biasing when initial conditions
are non-Gaussian. If we restrict ourselves to scales larger than those
relevant to galaxy formation, the galaxy density field can be thought
of as a local transformation of the density field smoothed over large
enough scales so that $\d \ll 1$, and thus expanded as (Fry \&
Gazta\~naga 1993)

\beq
\d_g = b_1 \d + \frac{b_2}{2} \d^2 + \ldots,
\label{locbias}
\eeq

\noindent which implies the following mapping for the power spectrum
and bispectrum

\beqa
P_g(k) &=& b_1^2 P(k) + b_1 b_2 \int d^3 q\ B^I(\k,\q),
\label{powerg} \\
B_g(k_1,k_2,k_3) &=& b_1^3 B_{123} + b_1^2 b_2 \Sigma_{123}^{I} + 
\frac{3}{2} b_1^2 b_2 \int d^3q\ T_{123}^I(\q),
\label{Bispg}
\eeqa 

\noindent where $T_{123}^I(\q)$ denotes the trispectrum
$T^I(\k_1,\k_2,\q)$ symmetrized over $\{k_1,k_2,k_3\}$, and
$\Sigma_{123}^{I}$ is defined in Eq.~(\ref{Q123}). The reduced
bispectrum then reads

\beq
Q_g= \frac{1}{b_1} Q_{123} + \frac{b_2^{\rm eff}}{b_1^2},
\label{QgNG}
\eeq
where the effective non-linear bias parameter is given by 

\beq b_2^{\rm eff} = b_2\ \left[ 1 + \frac{3}{2 \Sigma_{123}^{I}} \int
d^3q\ T_{123}^I(\q) -
\frac{Q_{123}^{I}}{\Sigma_{123}^{I}} \left( P_1^I \int d^3 q\
B^I(\k_2,\q)  + {\rm cyc.} \right)
\right] \label{b2eff} \eeq

\noindent The analogous result to Eq.~(\ref{QgNG}) for the 
case of the skewness was derived by Fry \& Scherrer (1994).  So far
the derivation does not assume any particular non-Gaussian model. In
order to compute the effective non-linear bias parameter, we need to
make assumptions about the initial bispectrum and trispectrum. The
particular form of the convolution integrals in Eq.~(\ref{b2eff})
require knowledge of the three- and four-point functions where two
points coincide. Using Eqs.(\ref{zetaI}-\ref{etaI}) we obtain

\beq
\int d^3 q\ B^I(\k_1,\q) = 2^{3/2} \sqrt{\xi^I(0)}\ P^I(k_1),
\eeq

\beq
\int d^3q\ T_{123}^I(\q) = 2^{3/2} \sqrt{\xi^I(0)}\ 
B_{123}^I + \frac{4}{3} \Sigma_{123}^I,
\eeq
so that

\beq 
b_2^{\rm eff} = b_2\ \left[ 3 - \sqrt{2 \xi^I(0)}\ Q_{123}^I \right]
\label{b2effchi}
\eeq

\noindent In this expression, the meaning of $\xi^I(0)$ is the
following. As we said above, the local bias model in
Eq.~(\ref{locbias}) holds for {\em smoothed} fields, so $\xi^I(0)$ is
the rms density fluctuation at the smoothing scale. The
smoothing filter is the Fourier transform, so in this case we should
replace $\xi^I(0) \approx \Delta^I(k)$ for the average scale $k
\approx (k_1+k_2+k_3)/3$ under consideration. Using Eq.~(\ref{delnl})
and Eq.~(\ref{QI}) we find

\beq
b_2^{\rm eff} \approx (3-\frac{8}{\sqrt{3}\pi})\ b_2 \approx 1.53\ b_2
\label{b2effchi2}
\eeq

\noindent Thus, for $\chi^2$ initial conditions, the usual Gaussian
initial conditions biasing formula is recovered (with no additional
scale or configuration dependence) provided a proper redefinition of
the non-linear bias parameter is made. Note that for other spectral
indices than $n=-1$, the resulting $b_2^{\rm eff}$ remains independent
of scale. In principle, there could be some residual dependence on
triangle configuration; however, for $n=-1.8$ the 2LPT results
described above give negligible residual configuration dependence as
well.

\section{Redshift Distortions}

In redshift space, the radial coordinate $\s$ of a galaxy is given by
its observed radial velocity, a combination of its Hubble flow plus
``distortions'' due to peculiar velocities. The mapping from
real-space position ${\bf \x}$ to redshift space is given by: 

\beq
\s=\x - f \ \uz(\x) {\hat z},
\label{zmap} 
\eeq 

\noindent where $f= d\ln D_1 / d\ln a \approx \Omega^{0.6}$, and
$\u(\x) \equiv - \v(\x)/({\cal H} f)$, where $\v(\x)$ is the peculiar
velocity field, and ${\cal H}(\tau) \equiv (1/a)(da/d\tau)= Ha$ is the
conformal Hubble parameter (with FRW scale factor $a(\tau)$ and
conformal time $\tau$).  In Eq.~(\ref{zmap}), we have assumed the
``plane-parallel'' approximation, so that the line-of-sight is taken
as a fixed direction, denoted by ${\hat z}$. Using this mapping, the
Fourier transform of the density field contrast in redshift space
reads (Scoccimarro et al. 1999)

\beq
\ds(\k) = \int  \frac{d^3x}{(2\pi)^3} {\rm e}^{-i \k\cdot\x}
{\rm e}^{i f k_z \uz(\x)} \Big[ \d(\x) + f \nabla_z \uz(\x) \Big].
\label{d_s}
\eeq

\noindent This equation describes the fully non-linear density field
in redshift space in the plane-parallel approximation.  In linear
perturbation theory, the exponential factor becomes unity, and we
recover the well known formula (Kaiser 1987)

\beq 
\ds(\k)=\d(\k)\ (1+f\mu^2)
\label{delta_sl},  
\eeq

\noindent where $\mu \equiv k_z/k$. Redshift distortions are trivial
to include for $Q^I_{123}$, since only linear PT is involved. Assuming
linear biasing, Eq.~(\ref{delta_sl}) gives the monopole of the reduced
bispectrum

\beq
Q^I_{s\ 123} =  \frac{\overline{ (1+\beta \mu_1^2)(1+\beta
\mu_2^2)(1+\beta \mu_3^2)}}{(1+2 \beta/3+\beta^2/5)^2} \times
\frac{Q^I_{123}}{b_1} \equiv A_s  \frac{Q^I_{123}}{b_1}, 
\label{Qs123}
\eeq

\noindent where $\beta \equiv f/b_1$, $\mu_i k_i \equiv \k_i \cdot
\hat{z}$ and the bar denotes angular average over triangle
orientations. Note that this results holds irrespective of the type of
non-Gaussian initial conditions. After some algebra, we obtain $A_s =
C_s/(1+2 \beta/3+\beta^2/5)^2$ with

\beq
C_s = 1+ \beta + \frac{2}{5}\beta^2 +\frac{2}{35}\beta^3 +
\frac{\beta^2 (7+3\beta)}{210} \left[
\frac{k_1^6+k_2^6+k_3^6-k_1^4k_2^2-k_1^2k_2^4-k_1^4k_3^2-k_2^4k_3^2-
k_1^2k_3^4-k_2^2k_3^4}{k_1^2 k_2^2 k_3^2} \right]
\label{As}
\eeq

\begin{figure}[t!]
\centering
\centerline{\epsfxsize=18truecm\epsfysize=18truecm\epsfbox{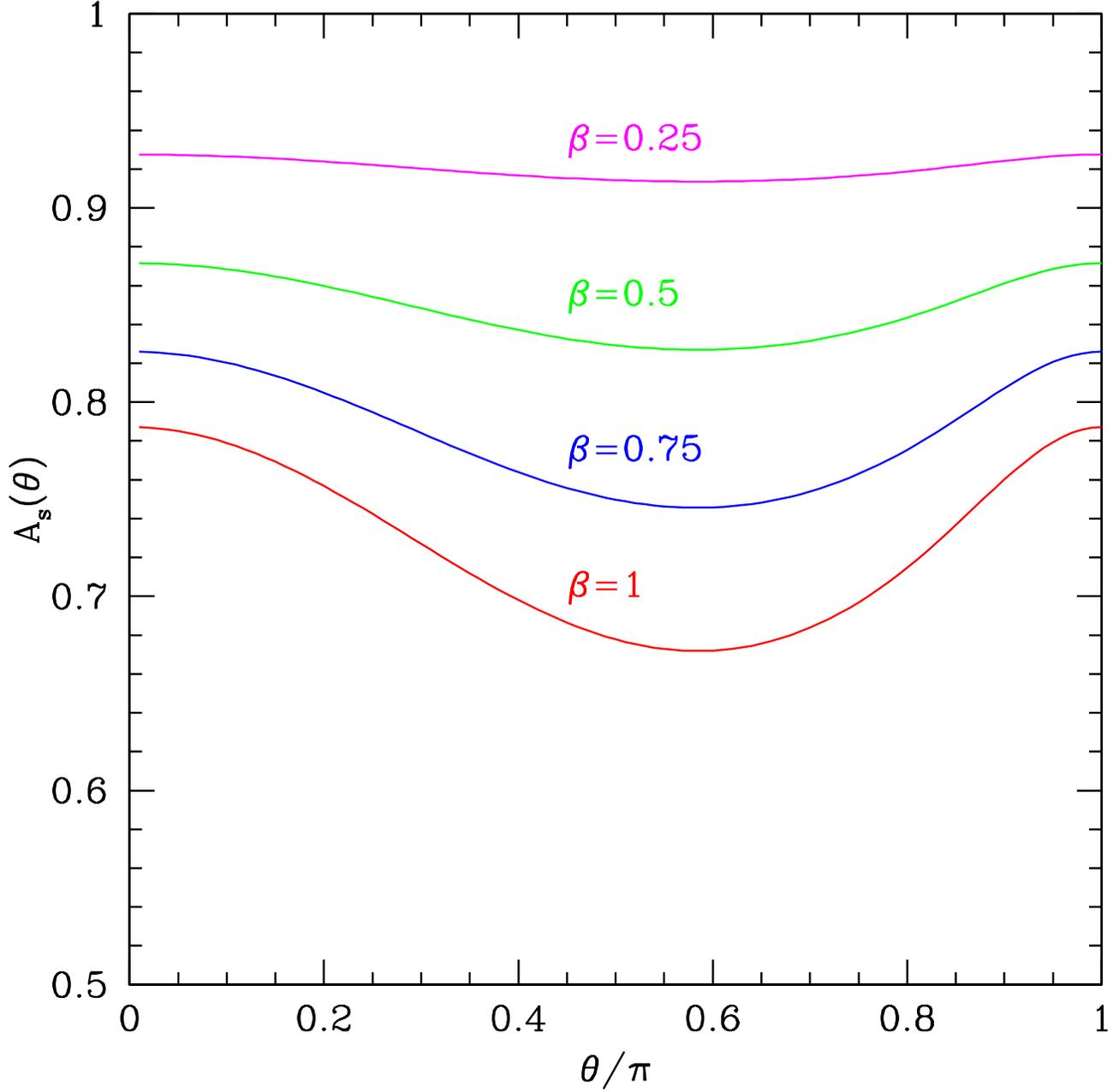}}
\caption{The correction factor for redshift distortions to the
primordial bispectrum, $A_s$, as a function of angle $\theta$ between
$\k_1$ and $\k_2$ for triangles with sides $k_1/k_2=2$, for different
values of $\beta \approx \Omega^{0.6}/b_1$.}
\label{figAs}
\end{figure}

In Fig.~\ref{figAs} we show the correction factor for redshift
distortions $A_s$. Unlike the Gaussian initial conditions case, where
the redshift-space correction is only about $10\%$ (Hivon et
al. 1995), for non-Gaussian models it can be significantly more,
depending on the value of $\beta$. The reason is simply due to the
scaling of the primordial bispectrum being dimensional rather than
hierarchical (as that generated by gravity from Gaussian initial
conditions). That is, from Eq.~(\ref{Qs123}) one expects that
approximately $A_s \approx 1/\sqrt{1+2/3 \beta + \beta^2 /5}$, which
describes very well the mean value of $A_s$ shown in
Fig.~\ref{figAs}. 

\begin{figure}[t!]
\centering
\centerline{\epsfxsize=18truecm\epsfysize=18truecm\epsfbox{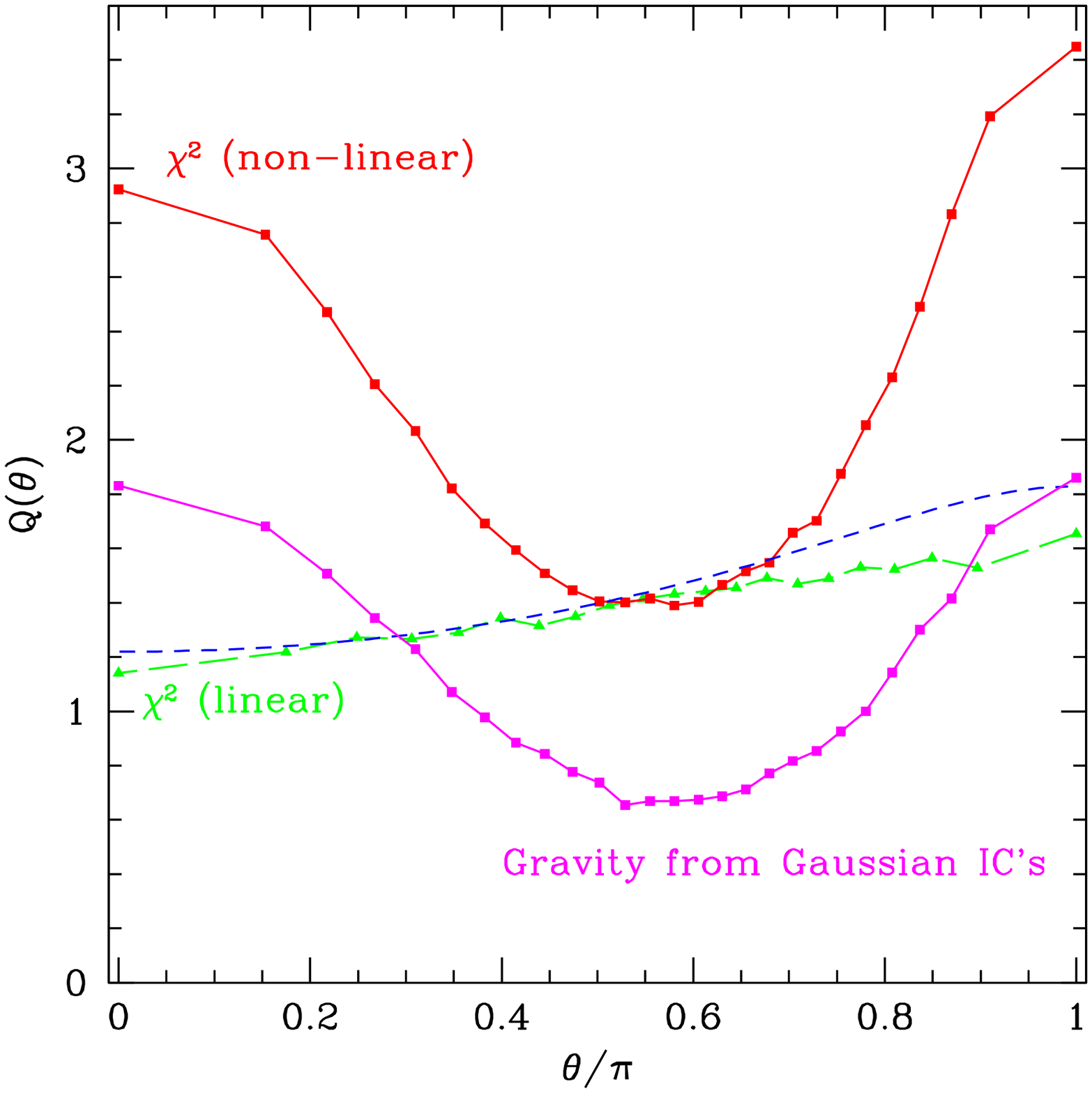}}
\caption{Same as Fig.~\protect\ref{figiso} but in redshift space.}
\label{figisoz}
\end{figure}

\begin{figure}[t!]
\centering
\centerline{\epsfxsize=18truecm\epsfysize=18truecm\epsfbox{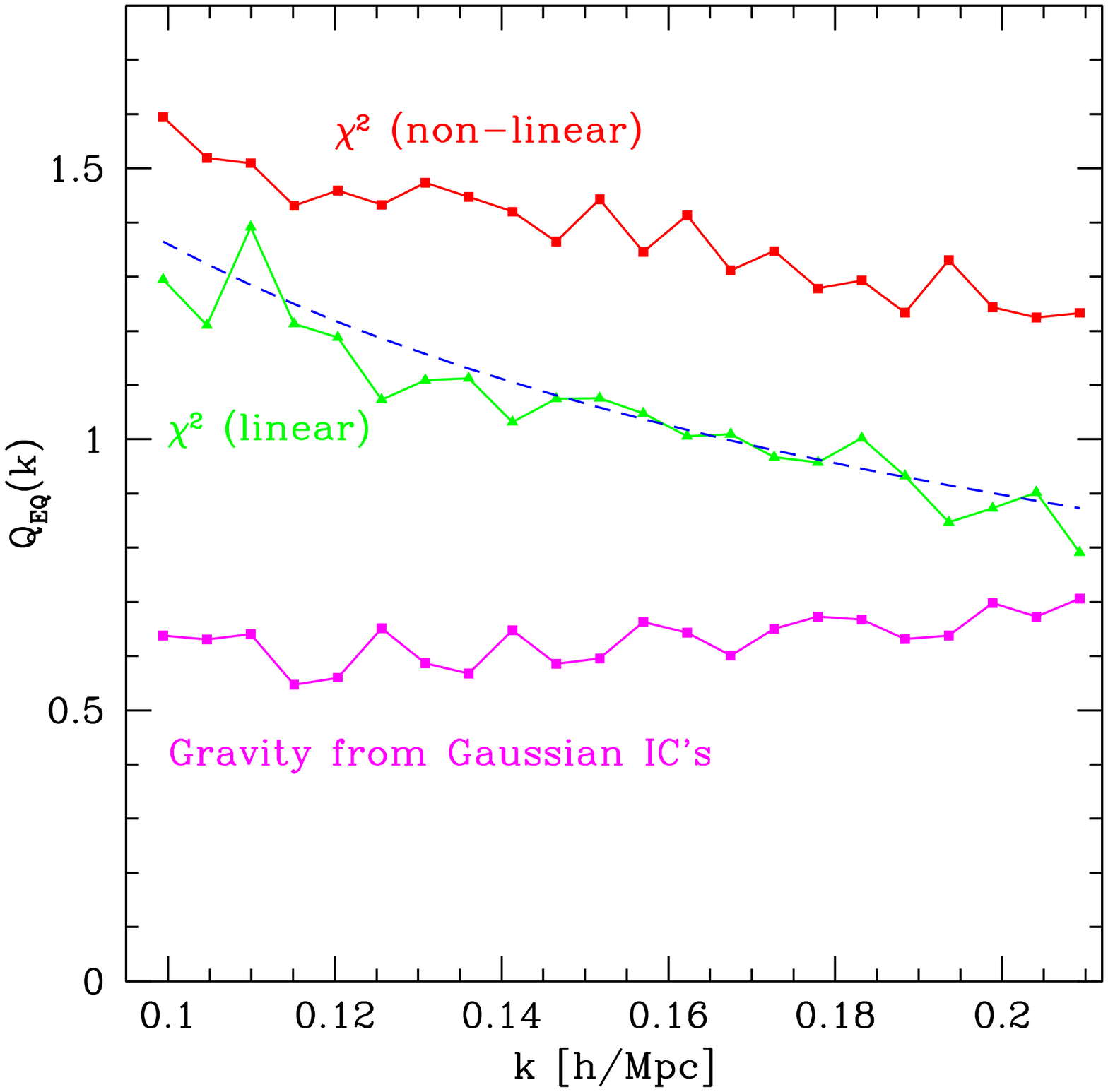}}
\caption{Same as Fig.~\protect\ref{figisoeq} but in redshift space.}
\label{figisoeqz}
\end{figure}

As seen from Eq.~(\ref{d_s}), the density field in redshift-space is
exponentially sensitive to the velocity field.  Thus, expanding this
effect by linear PT is only valid at very large scales.  In order to
incorporate the non-linear effects of the redshift-space mapping, we
ran 2LPT realizations in redshift space, which take into account this
mapping exactly. We assume a cosmological model with $\Omega=0.3$ and
$\Omega_\Lambda=0.7$, for which $\beta \approx
0.51$. Figures~\ref{figisoz} and~\ref{figisoeqz} show the results
corresponding to the same triangles as in Figs.~\ref{figiso}
and~\ref{figisoeq}, respectively.  The amplitudes in redshift space
are changed, but the overall behavior is the same: the $\chi^{2}$
model shows larger bispectrum amplitude and scale dependence than the
Gaussian initial conditions case.  Note that 2LPT calculations in the
latter case also automatically include non-linear effects besides
those due to redshift-space distortions (``loop corrections''), which
were not present in the predictions shown as dot-dashed lines in
Figs.~\ref{figiso} and~\ref{figisoeq}; this explains the mild scale
dependence of the equilateral bispectrum in Fig.~\ref{figisoeqz} {\em
opossite} to that of the $\chi^{2}$ model. In summary, we conclude
that the results of Section~3 are robust against the effects of
redshift distortions.

Another signature of primordial non-Gaussianity is provided by the
redshift-space power spectrum. As we discussed above, when initial
conditions are non-Gaussian, new couplings become available and
non-linear corrections become important at larger scales than in the
Gaussian initial conditions case. In redshift-space, this manifests
itself in a particular way, as we now discuss. Consider the
redshift-space power spectrum, from Eq.~(\ref{d_s}) we can write a
simple expression for the power spectrum in redshift space
(Scoccimarro et al. 1999)

\beq 
P_s(k,\mu) = \int \frac{d^3 r}{(2 \pi)^3} {\rm e}^{-i \k \cdot \r}
\Big\langle {\rm e}^{i f k \mu [ \uz(\x)-\uz(\x') ]} \Big[ \d(\x) + f
\nabla_z \uz(\x) \Big] \Big[ \d(\x') + f \nabla_z' \uz(\x') \Big]
\Big\rangle
\label{Ps},
\eeq

\noindent where $\r \equiv \x-\x'$. This is a fully non-linear
expression, no approximation has been made except for the
plane-parallel approximation. The factors in square brackets denote
the amplification of the power spectrum in redshift space due to
infall, and they constitute the only contribution in linear PT, giving
Kaiser's (1987) result 
\beq
P_s(k,\mu) = (1+ f \mu^2)^2 P(k)
\label{pkaiser}.
\eeq
\noindent The anisotropy of the power spectrum in redshift space can
be characterized by expanding it in multipole moments with respect to
$\mu$, the cosine of the angle between a given wave vector $\k$ and
the line of sight $\hat{z}$. This gives a monopole $P_0(k)=
(1+2/3f+f^2/5) P(k)$, and quadrupole $P_2(k)= (4/3f+4/7f^2)P(k)$. The
quadrupole to monopole ratio $P_2/P_0$ is thus sensitive to the matter
density parameter $\Omega$, in fact, assuming linear bias $P_2/P_0$ is
just a function of $\beta = f/b_1$ (Kaiser 1987; Hamilton 1992).
 
On the other hand, at smaller scales the pairwise velocity along the
line of sight in the exponential factor in Eq.~(\ref{Ps}) starts to
play a role. This eventually leads to a decrease in monopole and
quadrupole power with respect to the linear contribution; in
particular, the quadrupole changes sign and becomes negative. Note
that this effect is distinct from that of velocity dispersion
associated with clusters which takes place at smaller scales than
considered here, and leads to a strong negative quadrupole. In fact,
at the scales we work the decrease in the quadrupole to monopole ratio
can be understood from perturbative dynamics, as has been noted before
(Taylor \& Hamilton 1996; Fisher \& Nusser 1996; Hui, Kofman \&
Shandarin 1999; Scoccimarro et al. 1999).

From Eq.~(\ref{Ps}), one expects that the scale of quadrupole zero
crossing is very sensitive to the magnitude of infall velocities. In
redshift-space, infall into large-scale structures can become large
enough that ``near'' and ``far'' sides of large-scale structures in
real space reverse sides in redshift-space (Hui et al. 1999). At this
point statistical isotropy is recovered and thus the quadrupole
vanishes. The magnitude of infall at a given scale $k$ is essentially
given by the power spectrum $P_\theta(k)$ of the velocity divergence,
($\theta(\x) \equiv \nabla \cdot \u$), which evolves according to
(compare with Eq.\ref{pknl})

\begin{equation}
P_\theta(k) = P^I(k) + 2 \int d^3 q\ G_2(\k+\q,-\q) \ B^I(\k,\q),
\label{pknlt} 
\end{equation}

\noindent where the kernel $G_2(\k_1,\k_2) \equiv 3/7 + 1/2 \cos\theta
(k_1/k_2+k_2/k_1) + 4/7 \cos^2 \theta$ describes the second-order
evolution of the velocity field, similar to $F_2$ in Eq.~(\ref{den2})
for the density field. Thus, the velocity divergence power spectrum is
sensitive to the primordial bispectrum. For the $\chi^2$ model with
$n=-1$ model as in Section~3 gives (compare with Eq.\ref{delnl})

\begin{equation}
\Delta_\theta(k) = \left( \frac{k}{k_{nl}} \right)^2\ \left( 1 -
\frac{8}{7\sqrt{2} \pi} \frac{k}{k_{nl}} \right) \approx \left(
\frac{k}{k_{nl}} \right)^2\ \left( 1 - 0.26 \frac{k}{k_{nl}} \right).
\label{delnlt} 
\end{equation}

\noindent Thus, for $\chi^2$ initial conditions we expect large-scale
infall velocities to be smaller than for Gaussian initial
conditions. The effect of primordial non-Gaussianity on the pairwise
velocity distribution has been explored by Catelan \& Scherrer (1995).

\begin{figure}[t!]
\centering
\centerline{\epsfxsize=18truecm\epsfysize=18truecm\epsfbox{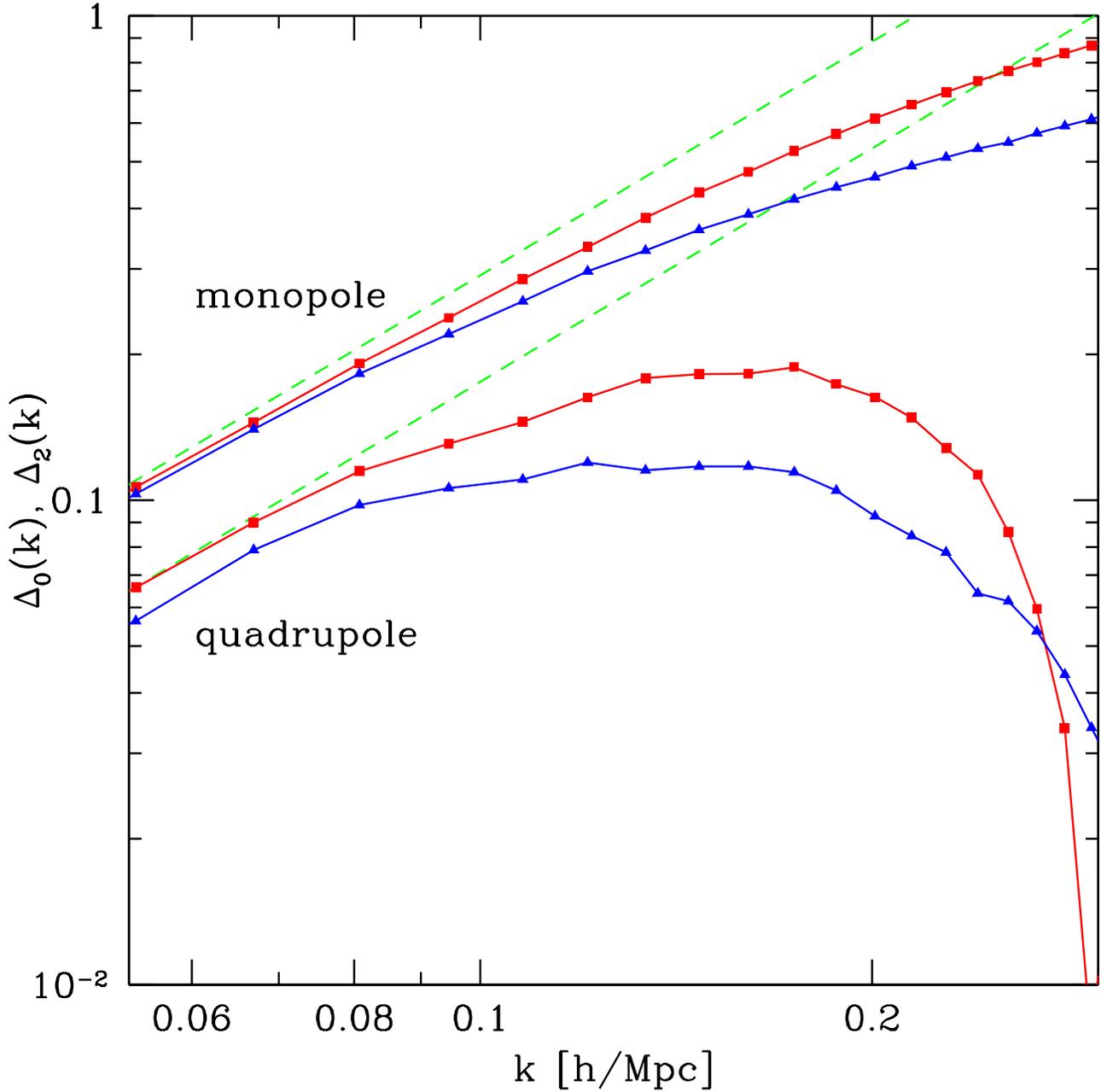}}
\caption{The power spectrum monopole (top set of curves) and
quadrupole (bottom) as a function of scale for scale-free initial
spectra $n=-1.4$ for Gaussian (squares) and $\chi^2$ initial
conditions (triangles). The dashed lines show the predictions of
linear PT.}
\label{figPk}
\end{figure}

\begin{figure}[t!]
\centering
\centerline{\epsfxsize=18truecm\epsfysize=18truecm\epsfbox{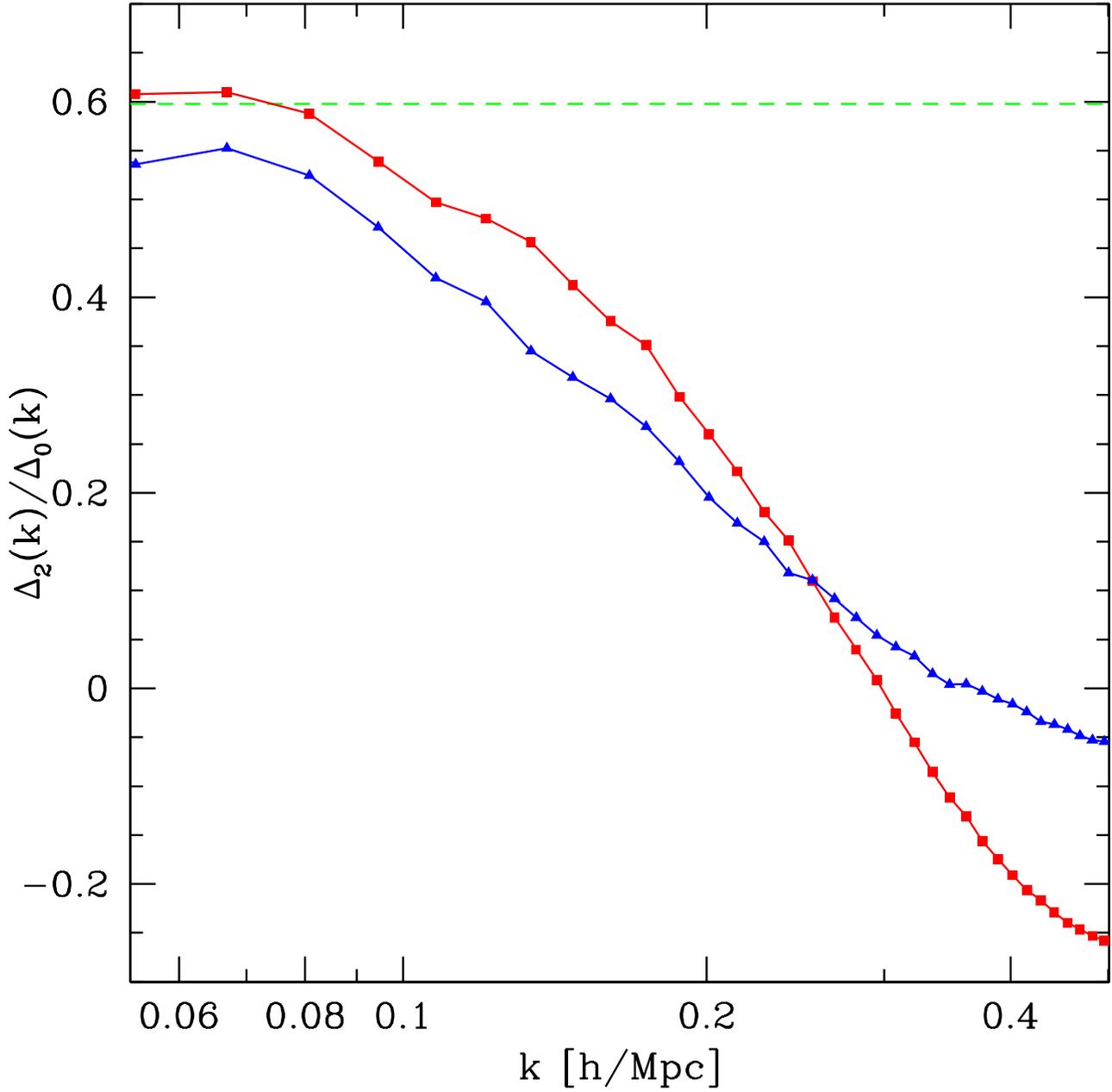}}
\caption{The power spectrum quadrupole to monopole ratio
$\Delta_2/\Delta_0$ as a function of scale for scale-free initial
spectra $n=-1.4$ for Gaussian (squares) and $\chi^2$ initial
conditions (triangles). The horizonal line denotes the large-scale
limit expected in linear perturbation theory.}
\label{figRP}
\end{figure}

Figures \ref{figPk} and \ref{figRP} show results of 100 2LPT
realizations of Gaussian and $\chi^2$ initial conditions with spectral
index $n=-1.4$, $k_{nl} = 0.33$ h/Mpc, $\Omega=0.3$ and
$\Omega_\Lambda=0.7$. The dashed lines show the predictions of linear
PT, Eq.~(\ref{pkaiser}). Clearly, non-linear effects are important
even at scales as large as $k = 0.06$ h/Mpc. Comparing the Gaussian
(squares) and the $\chi^2$ model (triangles), we see that the $\chi^2$
model shows a smaller amplification and the zero crossing of the
quadrupole happens at smaller scales, consistent with infall
velocities being smaller as described above. In fact, the quadrupole
has not yet reached its large-scale limit at $k=0.06$h/Mpc. We have
checked that this result is not an artifact by verifying that lowering
the power spectrum normalization brings the quadrupole to agreement
with linear PT. Therefore, the shape of the quadrupole to monopole
ratio is sensitive to primordial non-Gaussianity, in this case
becoming flatter than the corresponding Gaussian model with the same
power spectrum. At small scales sensitive to virial motions inside
clusters, however, we expect a different behavior. Models with
primordial positive skewness such as the $\chi^2$ model tend to have
more prominent ``fingers of god'' (Weinberg \& Cole 1992).

\section{Conclusions}

We have explored the predictions of $\chi^2$ initial conditions for
clustering statistics, in particular the power spectrum and
bispectrum. We found that extrapolation of the initial conditions
using linear perturbation theory is not accurate enough at the scales
best probed by current and future galaxy redshift surveys ($k \ga
0.05$ h/Mpc). This is not surprising, non-Gaussian initial conditions
provide additional non-linear couplings otherwise forbidden by
symmetry, thus non-linear gravitational corrections can become
important at larger scales than in the Gaussian case.

We showed that when non-linear corrections are included, the
bispectrum shows a similar configuration dependence as in the Gaussian
case; however its amplitude is much larger than the latter and has a
residual scale dependence not present in evolution from Gaussian
initial conditions. We included the effects of galaxy biasing and
showed that the galaxy bispectrum obeys the same relation to the dark
matter bispectrum as in the Gaussian case with a proper redefinition
of the non-linear bias parameter. Thus, a large linear bias decreases
the dependence of the bispectrum on triangle configuration, and a
non-linear bias contributes a constant independent of configuration
just as in the Gaussian case. The effects of redshift distortions were
shown to change the overall amplitude of Gaussian and non-Gaussian
bispectra, but clear difference between them remains.  Thus, we
conclude that the bispectrum is a very useful statistic to probe the
Gaussianity of initial conditions, at least for models where the
scaling is dimensional as in the $\chi^2$ case. Application of these
results to current galaxy surveys will be reported elsewhere
(Scoccimarro et al. 2000).

We also discussed the effects of non-Gaussianity on the power spectrum
in redshift space. The shape of the quadrupole to monopole ratio is
very sensitive to the large-scale pairwise velocity which depends on
the primordial bispectrum through non-linear corrections. For the
$\chi^2$ model, this leads to a smaller infall velocity and thus a
supression of overall power and smaller quadrupole zero crossing scale
compared to the case of Gaussian initial conditions. This provides an
additional signature of primordial non-Gaussianity that can be
explored with the future generation of galaxy redshift surveys.

\acknowledgments

I thank Francis Bernardeau, Josh Frieman, Enrique Gazta\~naga, Lam
Hui, Lev Kofman, and Jim Peebles for useful discussions.

\clearpage

\clearpage

\clearpage

\clearpage

\clearpage

\clearpage

\clearpage

\end{document}